# Temperature-Independent Logarithmic Current Slope in Interdiffusion-Broadened Silicon pn Junction toward Low-Power Thermometry

Sungjae Hong[1,2], Jeffrey A. Ivie[1], Tzu-Ming Lu[1,2], Christopher R. Allemang[1,2]

1)Sandia National Laboratories, Albuquerque, New Mexico 87123, United States

2)Center for Integrated Nanotechnologies, Sandia National Laboratories, Albuquerque, New Mexico 87123, United States

(*Electronic mail: sunhong@sandia.gov)

## Abstract

With the development of cryogenic technologies such as quantum computing, there is an increasing need for low-power cryogenic thermometers that minimize self-heating. Here, we report an anomalous temperature-independent logarithmic current slope, $d(\ln J)/dV$, in heavily doped silicon $p^+n^+$ junctions with significant dopant interdiffusion and highlight the advantages of our device as a low-power, cryogenic thermometer. Electrical characterizations from 2 to 420 K reveal that, at low forward bias, $d(\ln J)/dV$ remains nearly constant, which yields a voltage-temperature responsivity that is nearly independent of bias current. The observed transport cannot be fully understood with conventional diffusion or direct band-to-band tunneling. Instead, the results suggest the contribution of trap-assisted tunneling. These results establish interdiffusion-broadened silicon $p^+n^+$ junctions as a route to low-power temperature sensors while providing insight into carrier transport in heavily doped semiconductor junctions.

Quantum computing retains the promise of revolutionizing multiple fields, including economics, chemistry, material science,[1]among others but leading state-of-the-art designs in superconducting or semiconducting qubits require operations at cryogenic temperatures using dilution refrigerator technology.[2] Thermometry in this regime is demanding with sensors that must remain accurate and reliable while dissipating minimal power because self-heating can perturb qubit operations and fidelity.[3,4] Among various types of thermometers, silicon pn diodes are widely used owing to their on-chip integration compatibility, low cost, stability, and reasonable sensitivity.[5] However, they exhibit a nonlinear temperature response[6] and require relatively large bias currents to produce a measurable, stable voltage signal, leading to self-heating at cryogenic temperatures.[5]

These limitations mainly originate from the fundamental carrier transport mechanism in the silicon pn junction diode, diffusion-governed transport. In this regime the sensitivity, dV/dT, decreases with decreasing driving current,[5] so reliable readout requires a high bias current and, with it, significant self-heating. These limitations motivate the exploration of alternative junction systems whose carrier transport mechanism differ from the conventional diffusion-governed behavior.

Heavily doped $p^+n^+$ junctions have attracted considerable interest because they show transport mechanisms beyond conventional diffusion-driven behavior[7], finding applications in negative differential resistance (NDR) oscillators[8], amplifiers[9], and tunnel field-effect transistors[10,11]. Depending on doping concentration, depletion width, and defect states, $p^+n^+$ junctions can exhibit band-to-band tunneling (BTBT), trap-assisted tunneling (TAT), thermionic-field emission, or recombination currents.[7,9,12–16] Transport dominated by tunneling mechanism displays weak or negligible temperature dependence[7,12], whereas the forward bias current in conventional pn junctions follows an exponential dependence, leading to a strong temperature dependence of $d(\ln J)/dV$.[9] Although $p^+n^+$ junctions are generally expected to show BTBT and NDR, dopant interdiffusion and compensation can notably modify the electrostatic junction profile and carrier transport in the junction.[10,16,17] However, systematic studies on the transport physics in such compensated silicon $p^+n^+$ junction system are rare.[16]

In this study, we investigated the electrical transport characteristics of a compensated silicon $p^+n^+$ junction and explain it with an analytic model to highlight the advantages of our device as a cryogenic thermometer. The device exhibits a temperature-independent $d(\ln J)/dV$ in the forward bias regime of the diode current-voltage (J-V) characteristics, enabling current-independent thermometer sensitivity from 1 μA to 1 nA or less especially for cryogenic temperatures, which is significantly lower than typical operation current of silicon diode thermometer about 1-10 μA.[5] Therefore, our diode allows adjustment of operating current level depending on the temperature range of interest. Low current operation exhibits sub-nW power consumption, which minimizes the self-heating problem in existing silicon diode thermometers.[3,5]

With comprehensive analysis of the device, including secondary ion mass spectroscopy (SIMS), capacitance-voltage (C-V) characterization, and simulations with analytic transport model and Technology Computer Aided Design (TCAD)[16], it is shown that the temperature-independent slope originates from TAT-dominant carrier transport in the $p^+n^+$ junction. The BTBT current is suppressed by relatively broad depletion region with a depletion width of greater than 20 nm in the $p^+n^+$ junction. Consequently, carrier transport becomes TAT-dominant, giving rise to a nearly temperature-independent logarithmic slope.

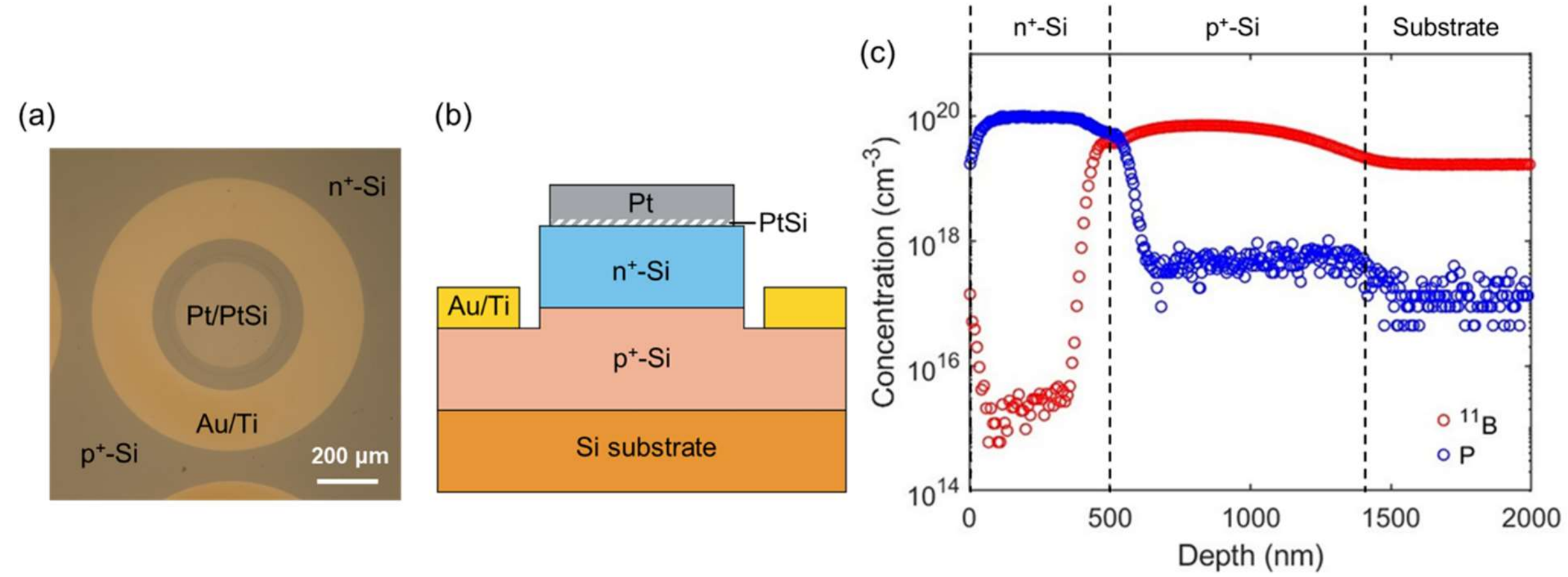

**Figure 1. Structure and depth profile of the p$^+$n$^+$ diode. (a) Optical microscope image of the diode. (b) Cross-sectional schematics of the device structure. (c) SIMS doping profile.**

In Figure 1a and 1b, the optical microscope image and the schematic cross-sectional structure of the p$^+$n$^+$ junction diode are presented. The device was fabricated on epitaxial silicon layers grown by chemical vapor deposition (CVD) at Lawrence Semiconductor Research Laboratory (LSRL) on a boron-doped (100) silicon wafer. Specifically, a 1 µm thick boron-doped silicon and a 500 nm thick phosphorous-doped silicon layer were sequentially deposited on the starting wafer. Following the epitaxial growth, the mesa regions with four different diameters were defined by photolithography and etched using $SF_6$, targeting an 800 nm etch depth. The mesa etch extends into the p$^+$-silicon layer, resulting in the p$^+$n$^+$ junction being confined to the mesa area. To form a contact to the n$^+$-silicon, 100 nm platinum was deposited by electron-beam (E-beam) evaporation and subsequently annealed at 250 °C for 30 minutes using a rapid thermal annealing system to form platinum silicide. Finally, a 5 nm Ti / 100 nm Au stack was deposited by E-beam evaporation as the contact for the p$^+$-silicon.

SIMS analysis in Figure 1c reveals the clear doping profile across the $p^+n^+$ junction. The junction is located at a depth of approximately 500 nm, with a spatial overlap of approximately 100 nm between the boron and phosphorus concentration profiles at the junction. This overlap indicates interdiffusion of the p- and n-type dopants during the growth process.[18] The SIMS results are corroborated by spreading resistance profiling (SRP) data, which show the same ~500 nm junction depth and a gradual variation in carrier concentration across the junction (Supporting Information Figure S1). The SRP indicates the maximum active carrier concentration in $n^+$-Si region about $2*10^{-19}$ $cm^{-3}$, which is lower than the dopant concentration. It possibly implies that the dopant activation rate is low in $n^+$-Si region.

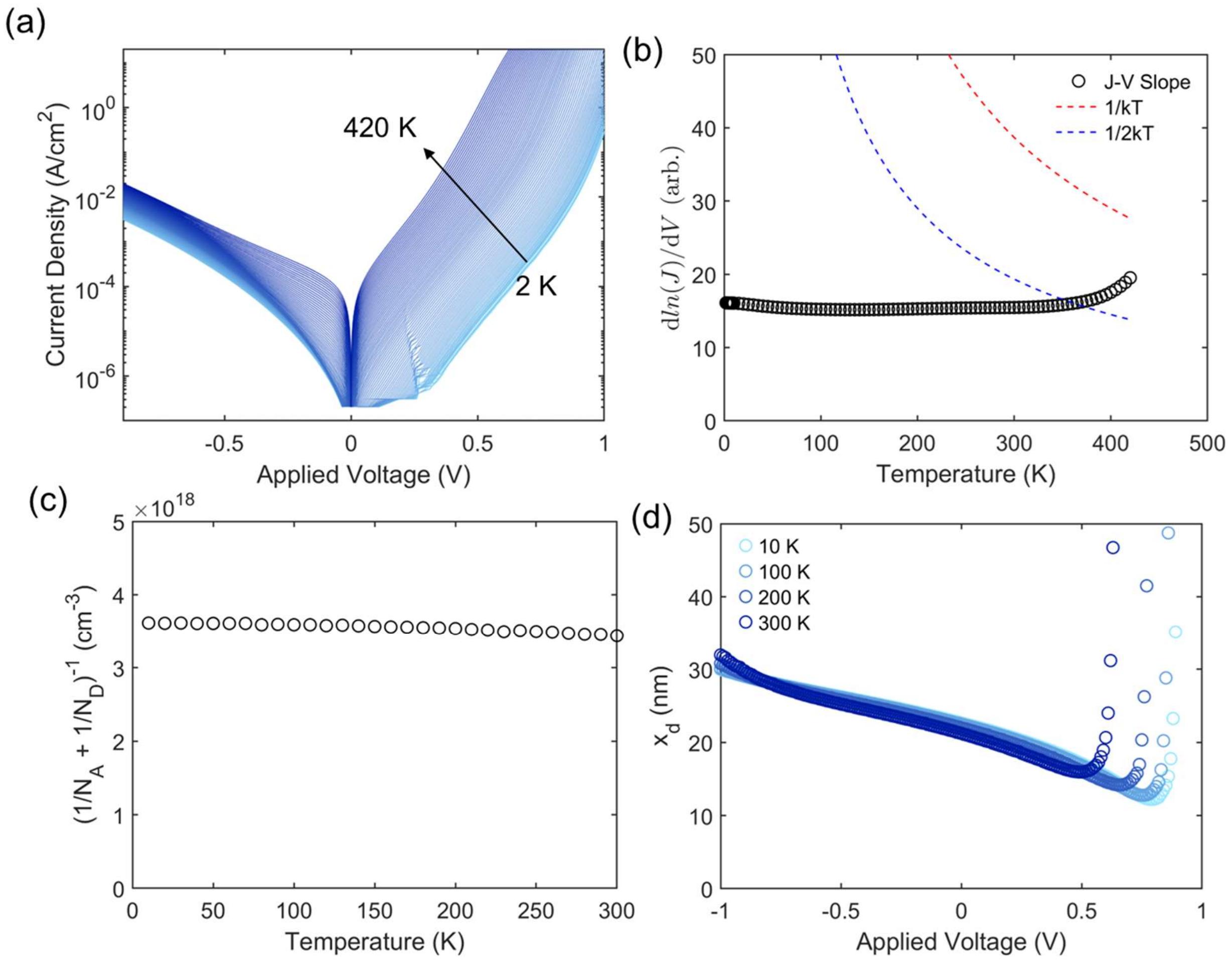

**Figure 2. Electrical characterization of a silicon p$^+$n$^+$ diode. (a) Temperature-dependent J-V characteristics from 2 to 420 K. (b) Log-slope vs temperature extracted from J-V curves. The dashed lines represent the expected slope from 1/kT and 1/2kT dependence. (c) Effective carrier concentration vs temperature obtained from C-V characteristics. (d) Depletion thickness at various temperatures.**

After fabrication, the devices were characterized by temperature-dependent J-V and C-V measurements in a cryogen-free closed-cycle cryostat over 2 K to 420 K. The J-V characteristics were measured with a Keithley 2400 source meter and C-V characteristics were measured using an arbitrary waveform generator, current preamplifier, and Stanford SR865A lock-in amplifier.

Figure 2a presents the measured J-V characteristics of a diode with a 250 μm diameter. The device shows rectifying behavior but without the NDR that would be expected of BTBT under forward bias. The log slope, $d(\ln J)/dV$, in the weak forward bias regime was obtained by linear fitting to $\ln J$ vs V under moderate current injection condition of $1*10^{-4}$ A/cm$^2$ < J < $1*10^{-3}$ A/cm$^2$, and is plotted as a function of temperature in Figure 2b. Notably, this slope remains nearly constant over the entire temperature range, showing only minimal variation at low and high temperatures. In contrast, the slope in a conventional pn diode decreases as temperature increases when diffusion or Shockley-Read-Hall (SRH) generation-recombination currents dominate the overall transport as the current density follows an exponential dependence on $qV/kT$ or $qV/2kT$ (Figure 2b, dashed lines), respectively, where $q$, $V$, $k$, and $T$ are the elementary charge, bias voltage, Boltzmann constant, and temperature, respectively.[15] Measurements obtained from devices with different diameters show the current scaling well with device area (Supporting Information Figure S2), which suggests the overall characteristics are dominated by the conduction through the bulk, not through the sidewall. All these devices exhibit similar nearly constant $d(\ln J)/dV$ regardless of the device size (Supporting Information Figure S3).

C-V measurements were also performed for the device with 900 μm diameter (Supporting Information Figure S4). The capacitance decreases as the magnitude of the reverse bias increases (Figure S4, inset) and the $1/C^2$ -V curves show a linear dependence as the bias voltage increases. From the $1/C^2$ vs V curves, the carrier concentration and the depletion width were extracted as functions of temperature and bias voltage, respectively (Figure 2c and d), using the following equations:[15]

$$N = \frac{-2}{q\varepsilon_r\varepsilon_0[d(1/C^2)/dV]}$$

$$C = \sqrt{\frac{q\varepsilon_r\varepsilon_0 N}{2(\phi_i - V)}} = \frac{\varepsilon_r\varepsilon_0}{x_d}$$

where N, $\varepsilon_r$, $\varepsilon_0$, C, $\phi_i$, $x_d$ are carrier concentration, relative permittivity, vacuum permittivity, capacitance, built-in potential, and depletion width, respectively. Here, the carrier concentration is extracted by fitting to the linear part in $1/C^2$ -V curves. The carrier concentration is nearly temperature-independent, with a value of $3.6 \times 10^{18}$ $cm^{-3}$. The temperature-independence indicates that the free carriers mainly originate from degenerately doped impurities.[15] These carrier concentrations are much lower than the dopant concentration in the $p^+$ and $n^+$ regions from SIMS, but can be explained by the dopant compensation in the junction due to significant dopant interdiffusion. The extracted depletion width is 23 nm at zero bias and increases to 30 nm with reverse bias of -1 V. These values are larger than ~ 10 nm scale reported for some abrupt, degenerately doped Si $p^+n^+$ tunnel junctions [9,19], consistent with the reduced carrier concentration.[17] This broadened depletion region significantly decreases the BTBT probability,[20,21] thereby suppressing NDR.

To identify the underlying transport mechanism, the J-V characteristics were analyzed using a fit to an analytical diode model. The forward bias current was modeled using a two-diode equation consisting of two components with distinct temperature dependences, which is a similar approach to the study by J. Jang, et al.[22]:

$$J(T,V) \approx J_1(T)\left[\exp\left(\frac{qV}{kT}\right) - 1\right] + J_2(T)\left[\exp\left(\frac{V}{E_2}\right) - 1\right]$$

where $J(T,V)$ is the total current density, $J_1(T)$ and $J_2(T)$ are the saturation current densities associated with each component and $E_2$ is the effective activation energy for the second component. Here, the first term represents the conventional diffusion current, whereas the second term accounts for other transport mechanisms combined, including TAT and SRH generation-recombination currents, while BTBT is ruled out here by the lack of forward bias NDR. The effects of series resistance was neglected because the model was intended to describe the moderately forward-biased regime. The fitting was performed based on the Levenberg-Marquardt nonlinear least square algorithm.

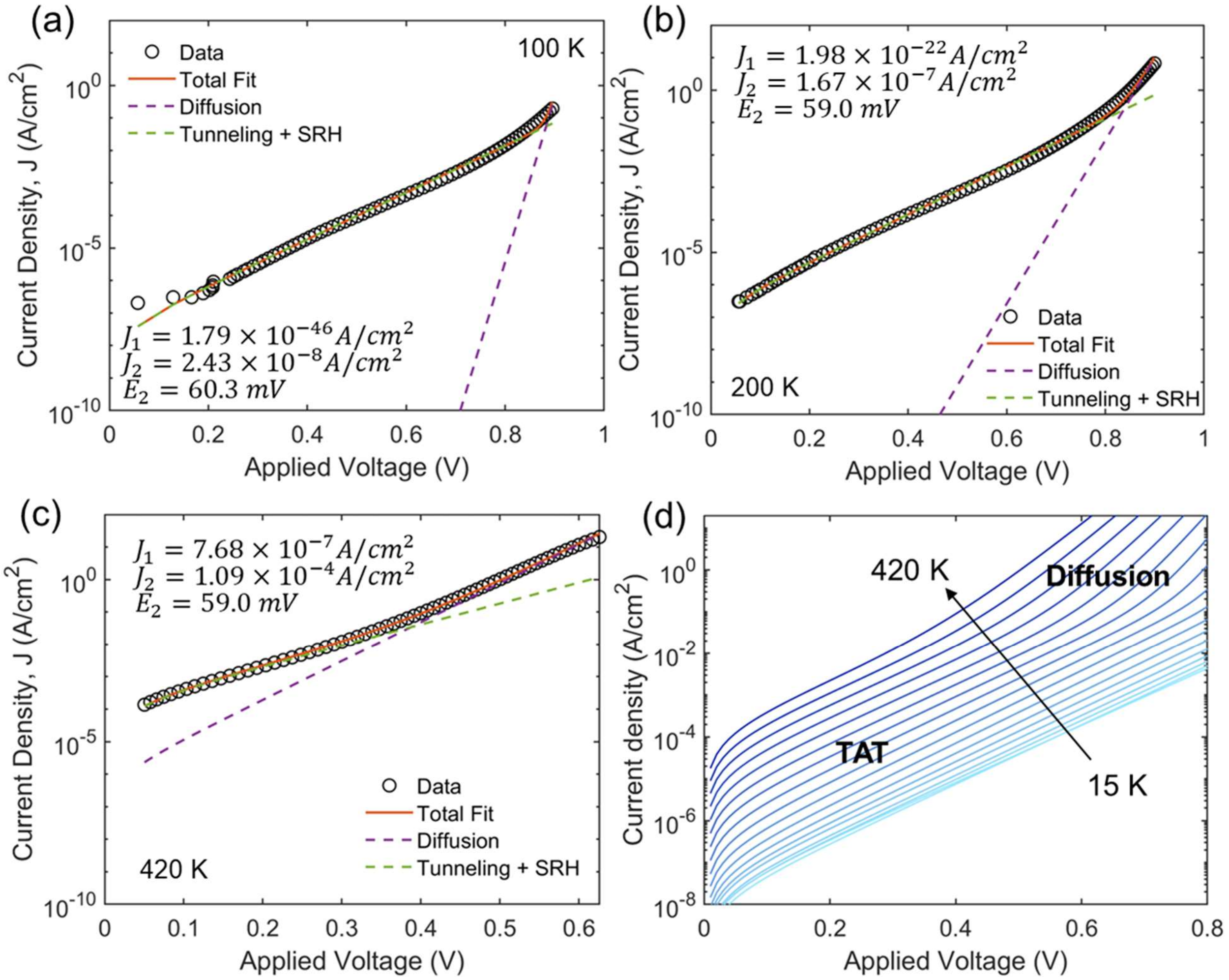

**Figure 3 J-V characteristics in forward bias regime at representative temperatures and the fitting results. (a) 100 K, (b) 200 K, (c) 420 K. The fitted parameter values are presented in each plot. (d) Simulated J-V characteristics from 15 to 420 K.**

Figure 3a-c show representative fitting results at 100, 200, and 420 K. First, the result at 100 K indicates that the non-diffusion contribution dominates the forward bias current throughout the measured bias voltage. The data are well described by the second term alone. As temperature increases, the fitting result at 200 K shows the diffusion current makes a noticeable contribution at high forward bias above 0.8 V, while the low-bias regime is still dominated by other mechanisms. At 420 K, diffusion current becomes dominant from moderate forward bias

above ~ 0.4 V. Figure 3d summarized the simulated J-V characteristics, where we can distinguish the diffusion and non-diffusion dominated regimes.

Figure 4 presents more detailed analysis of the simulation results. Figure 4a shows the $J_1(T)/T^3$ vs $1/kT$ relation. The slope in the plot represents the bandgap of the semiconductor comprising the pn junction according to the following relations for the diffusion current in pn junction:[15]

$$J_1(T), \propto q\left(\frac{D_n}{N_A L_n} + \frac{D_p}{N_D L_p}\right) n_i^2$$

$$n_i^2 \propto T^3 exp(-E_g/kT)$$

where $D_n$, $D_p$, $N_A$, $N_D$, $L_n$, $L_p$, $n_i$, and $E_g$ represent the electron and hole diffusion coefficient, acceptor and donor doping concentration, the diffusion lengths in the n and p materials, intrinsic carrier concentration, and bandgap, respectively. The extracted $E_g$ of 1.116 eV is close to the bandgap of silicon,[15] indicating that the fitting successfully captures the diffusion. In addition, extracted $E_2$ values are plotted as a function of temperature in Figure 4b. The fitting reliability analysis is discussed in Supporting Information Figure S5 and Note 1. The $E_2$ values in Figure 4b remain nearly temperature-independent. The extracted $E_2$ values deviate significantly from $2kT$ trend (Figure 4b, dashed line) which is expected from SRH-related transport.[9,15] This observation suggests that the behavior is more consistent with tunneling transport than thermally activated transport such as SRH current. In contrast to the temperature-independent $E_2$, the prefactor $J_2(T)$ shows a clear temperature dependence (Figure 4c). This implies that the carrier transport is affected by a thermally-activated carrier population or occupation statistics of tunneling states, as expected for TAT.[13,23] The $J_2(T)$ trend in Arrhenius plot does not show clear linearity, which means that the mechanism does not follow simple SRH-type behavior (Supporting Information Figure S6). This observation further supports that the device's charge transport is not dominated by SRH mechanisms for low to medium forward bias voltages. On the other hand, the temperature dependence of $J_2(T)$ is well described by an effective expression motivated by multiphonon-involved TAT. Starting from the phonon-weighted transition rate formalism of Makram-Ebeid and Lannoo[24], and assuming a single vibrational mode and an approximately linear energy dependence of the logarithm of the electronic tunneling rate over the dominant transition paths, we obtain the following expression for $J_2(T)$,

$$J_2(T) = J_{20} \exp\left[B\left(\coth\frac{\hbar\omega}{2kT} - 1\right)\right]$$

, where $J_{20}$ is the low-temperature limiting value, $B$ is a dimensionless parameter combining electron-phonon coupling and the energy sensitivity of tunneling, and $\hbar\omega$ is an effective vibrational energy. The red dashed line in Figure 4c shows the fit to this expression. The extracted vibrational mode energy of 29.2 mV lies within the energy range of acoustic phonons in Si; however, this effective energy alone does not identify the participating phonon modes. The derivation and underlying assumptions are provided in Supporting Information Note 2.

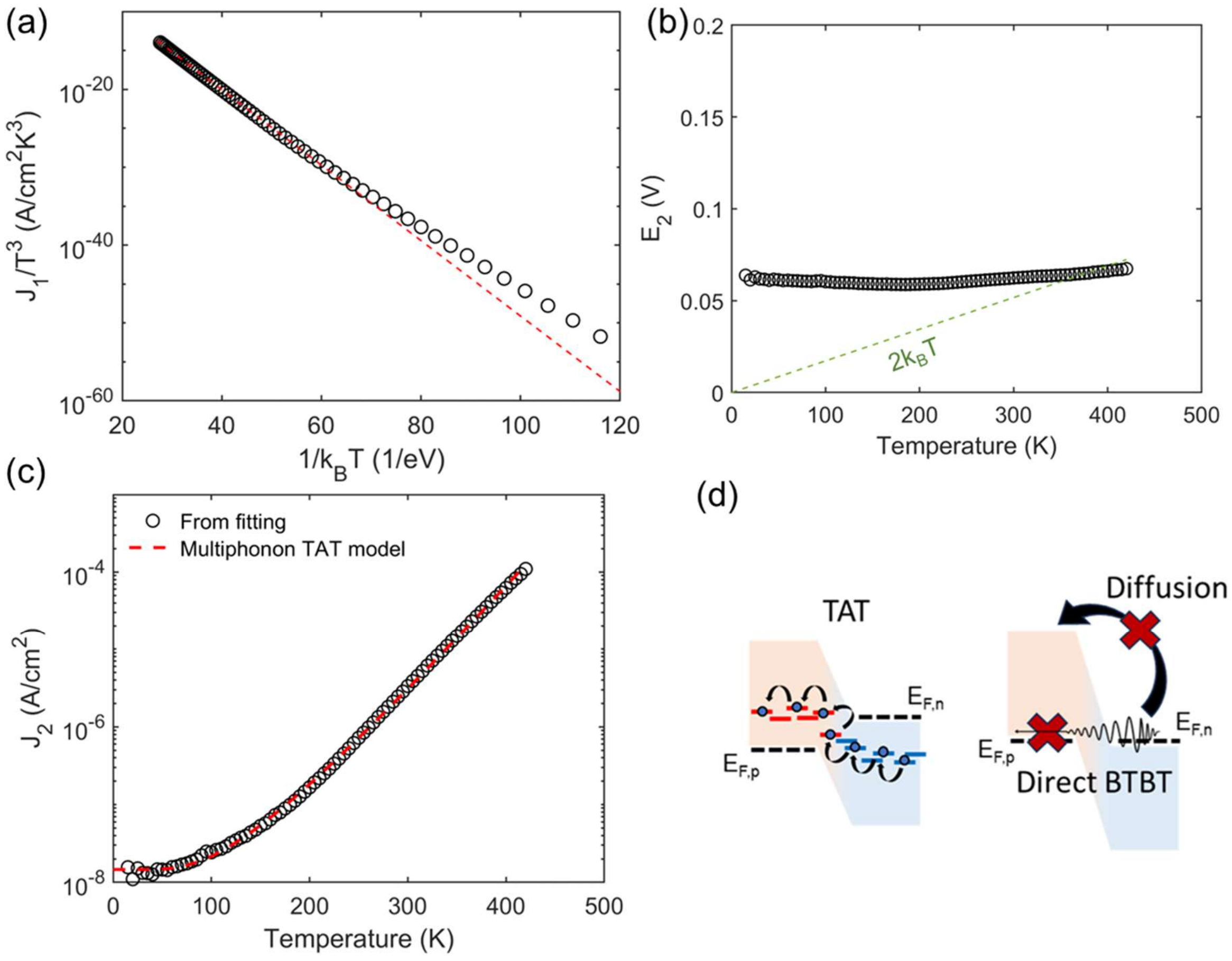

**Figure 4. Fitted parameters from the analytic diode model at various temperatures, (a) $J_1$, (b) $E_2$, (c) $J_2$. (d) Transport mechanism in the Si $p^+n^+$ junction diode with widened depletion width.**

Figure 4d summarizes the proposed charge transport mechanism in the $p^+n^+$ junction diode. In the device, diffusion, SRH recombination-generation, and BTBT currents are substantially suppressed. First, the diffusion contribution is small because the large built-in potential limits carrier injection until sufficiently high forward bias is applied. Secondly, the current is dominated by TAT due to the strong electric-field enhancement of the tunneling process under low to moderate forward bias voltages, whereas the SRH contribution becomes relatively small due to the limited carrier injection under low bias. More intriguingly, BTBT current, which is common in $p^+n^+$ junction diodes, is not significantly observed in our device. This may be attributed to the significant interdiffusion of dopant as shown in the previous section. In addition, at dopant concentrations approaching $10^{20}$ – $10^{21}$ $cm^{-3}$, isolated impurity states are expected to overlap, resulting in the formation of impurity bands. Furthermore, dopant clustering, including boron interstitial clusters and phosphorus-vacancy complexes, have been predicted and experimentally observed in degenerately doped silicon.[25] Such clustered defects can introduce deep localized electronic states inside the bandgap and may provide intermediate states for TAT. Taken together, these observations suggest that TAT is the dominant transport mechanism in the present device and causes the temperature independent $d(\ln J)/dV$. Similar transport has been also reported in a study of a Ag/p-GaN/In Schottky diode and was explained with field-emission mediated by a large amount of defect states near the GaN surface or deep level defect band induced by high doping.[26] More detailed study of the transport mechanisms through physics-based TCAD simulations can be found in a companion paper.[16]

A nearly temperature-independent $d(\ln J)/dV$ is rarely observed in Si pn junction diodes, making it attractive for thermometer applications since it enables nearly constant temperature sensitivity over a wide range of temperature. Figure 5a and 5b show the voltage-temperature characteristics and temperature sensitivity dV/dT, respectively, of the 250 μm diameter device at various currents from 1 nA to 10 μA. In Figure 5a, the curves have similar slopes throughout the temperature range except for at low driving current where the voltage approaches zero as temperature increases. As a result, we can observe the sensitivity curves in Figure 5b overlap, particularly at low temperatures where the tunneling current dominates the overall conduction. Above 200 K, the sensitivity curves remain nearly identical at high driving currents but gradually decrease as the driving current decreases because the bias voltages for the low currents approach zero as temperature increases. The current-independent sensitivity originates from the temperature-independent $E_2$ in the device. Assuming that tunneling current dominates the transport and the bias is sufficiently high to ignore the saturation-current term, the current through the device can be simplified as

$$J \approx J_2(\mathrm{T})\exp\left(\frac{V}{E_2}\right)$$

Then, by rearranging the equation, the voltage at a current J and temperature T can be expressed as

$$V(J,T) = E_2 ln(J) - E_2 ln(J_2(T))$$

Here, $E_2$ is temperature-independent in our device and J is fixed during the measurement of thermometer operation. Therefore, the sensitivity dV/dT now becomes

$$\frac{dV}{dT} = -E_2 \frac{1}{J_2}\frac{dJ_2}{dT}$$

, which is independent of the driving current J. The sensitivity calculated from fitting results is plotted as red cross in Figure 2b. The difference between data and calculation comes from small temperature dependence of $E_2$ and contribution from the diffusion current.

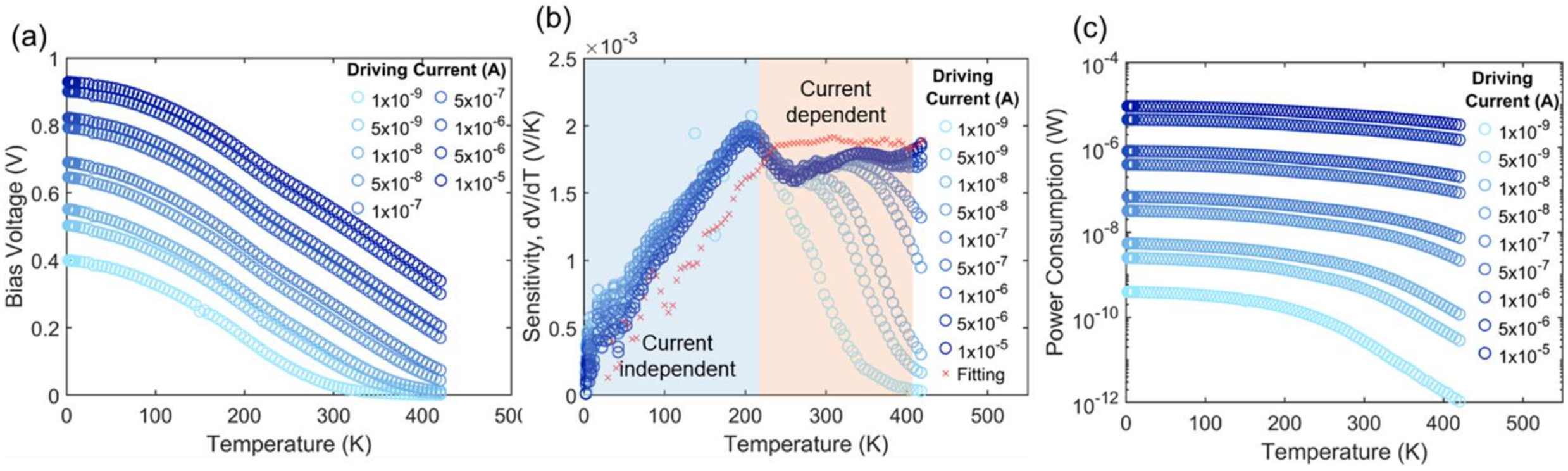

**Figure 5. Temperature dependent thermometer characteristics at various driving currents. (a) Voltage across the diode. (b) Thermometer sensitivity dV/dT. (c) Static power consumption.**

This current-independent sensitivity offers significant advantages for cryogenic temperature sensing. At relatively high temperatures, where self-heating is negligible, the device can be driven at high current for high sensitivity and stability. Conversely, at cryogenic temperatures, the device can be operated at very low current as the sensitivity is essentially independent of driving current. In that case, the power consumption can be substantially reduced, thereby minimizing the self-heating-induced temperature instability. Figure 5c presents the power consumption in the 250 μm diameter device as a function of temperature at various driving current. The power consumption decreases to 400 pW as the driving current goes down to 1 nA. Therefore, the device is well suited for temperature sensing over a wide range of temperatures, including cryogenic temperatures. One limitation of the present device is the reduced sensitivity at extremely low temperature, where the sensitivity approaches zero.

In summary, we investigated the electrical transport in a heavily doped $p^+n^+$ junction diode with a relatively wide depletion region caused by dopant interdiffusion. The device exhibits an unusual temperature-independent $d(\ln J)/dV$ under forward bias over a wide temperature range. Comprehensive characterizations, and analytical transport modeling revealed that the observed behavior stems from tunneling-dominant transport. Significant dopant interdiffusion near the junction results in reduced carrier concentration, widening the depletion region to more than 20 nm. The large depletion region suppresses BTBT, while the large built-in potential suppresses diffusion current, leading to TAT-dominated transport. More detailed analysis can be found in a modeling study by Juan P. Mendez, et al.[16]

The temperature-independent $d(\ln J)/dV$ provides nearly current-independent thermometer sensitivity, particularly at cryogenic temperatures. Consequently, the device can operate at substantially lower current levels than conventional silicon diode thermometers while maintaining nearly the same sensitivity over a wide range of driving currents. The resulting sub-nW power consumption significantly reduces self-heating, making the device attractive for low-power cryogenic temperature sensing applications. These findings offer insight into carrier transport in compensated heavily doped silicon $p^+n^+$ junctions and demonstrate their potential for practical cryogenic thermometry.

## Acknowledgement

The authors thank Evan M. Anderson, Andrew J. Leenheer, and Xujiao Gao for helpful conversations. This work was supported by the U.S. Department of Energy (DOE) Advanced Materials and Manufacturing Technologies Office (AMMTO) project BEATS. Sandia National Laboratories is a multimission laboratory managed and operated by National Technology and Engineering Solutions of Sandia, LLC., a wholly owned subsidiary of Honeywell International, Inc., for the U.S. Department of Energy's National Nuclear Security Administration under contract DE-NA-0003525. Work was performed, in part, at the Center for Integrated Nanotechnologies, a U.S. DOE, Office of Basic Energy Sciences user facility. This paper describes objective technical results and analysis. Any subjective views or opinions that might be expressed in the paper do not necessarily represent the views of the U.S. Department of Energy or the United States Government. This article has been authored by an employee of National Technology \& Engineering Solutions of Sandia, LLC under Contract No. DE-NA0003525 with the U.S. Department of Energy (DOE). The employee owns all right, title and interest in and to the article and is solely responsible for its contents. The United States Government retains and the publisher, by accepting the article for publication, acknowledges that the United States Government retains a non-exclusive, paid-up, irrevocable, world-wide license to publish or reproduce the published form of this article or allow others to do so, for United States Government purposes. The DOE will provide public access to these results of federally sponsored research in accordance with the DOE Public Access Plan https://www.energy.gov/downloads/doe-public-access-plan.

## AUTHOR DECLARATIONS

### Conflict of Interest

The authors have no conflicts to disclose.

**Author contribution**: Sungjae Hong: Conceptualization (equal); Data curation (equal); Formal analysis (lead); Investigation (equal); Methodology (equal); Software (lead); Validation (equal); Visualization (lead); Writing - original draft (lead); Writing - review & editing (equal). Jeffrey A. Ivie: Conceptualization (supporting); Data curation (equal); Funding acquisition (equal); Project administration (supporting); Supervision (supporting); Writing - review & editing (supporting). Tzu-Ming Lu: Conceptualization (equal); Funding acquisition (equal); Investigation (equal); Methodology (equal); Project administration (equal); Resources (lead); Supervision (equal); Writing - original draft (supporting); Writing - review & editing (supporting). Christopher R. Allemang: Funding acquisition (equal); Project administration (lead); Writing - original draft (supporting); Writing - review & editing (supporting).

## DATA AVAILABILITY STATEMENT

The data that support the findings of this study are available from the corresponding author upon reasonable request.

**Supporting Information**

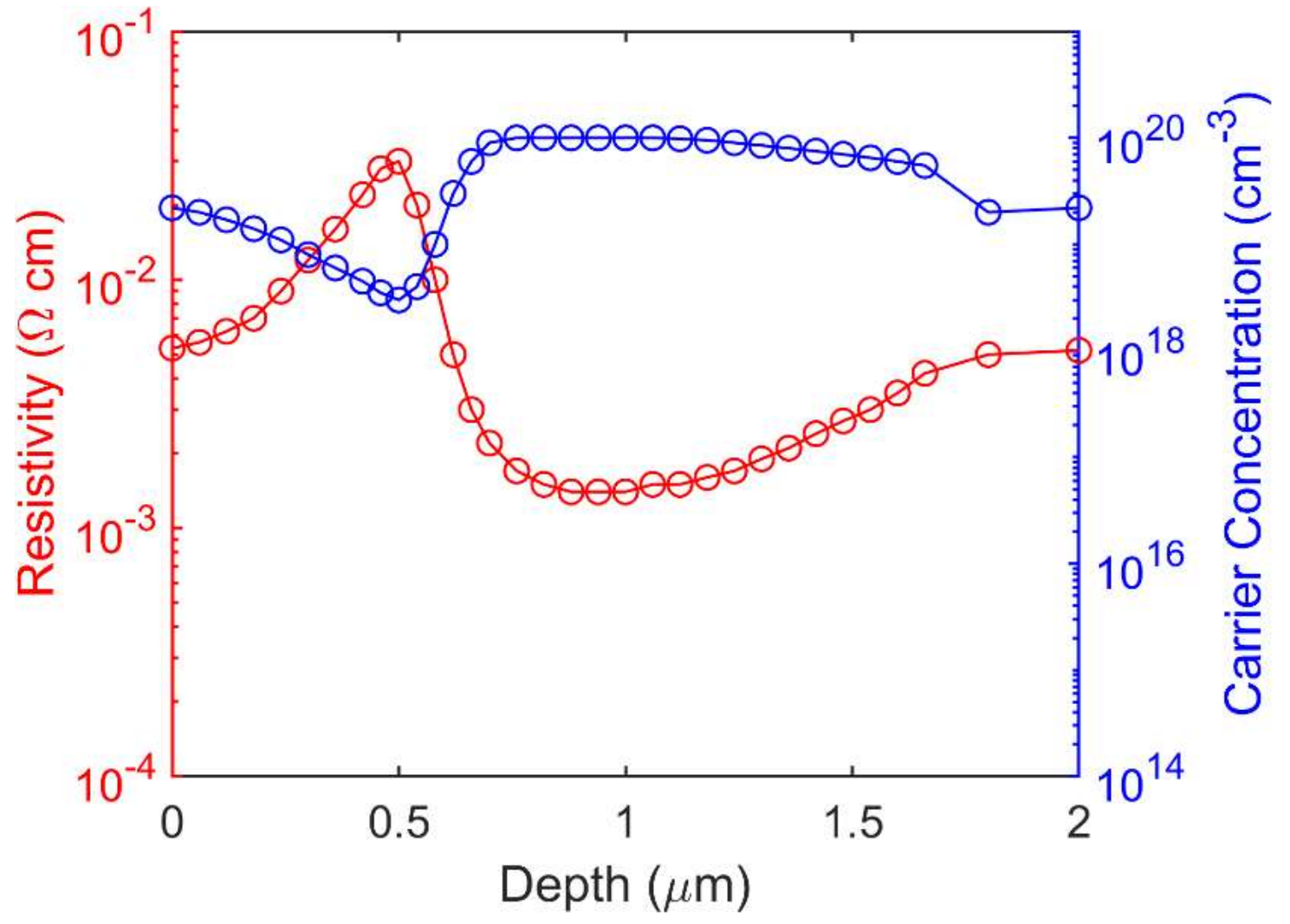

**Figure S1 Spreading Resistance Profile of the heavily-doped silicon $p^+n^+$ junction.**

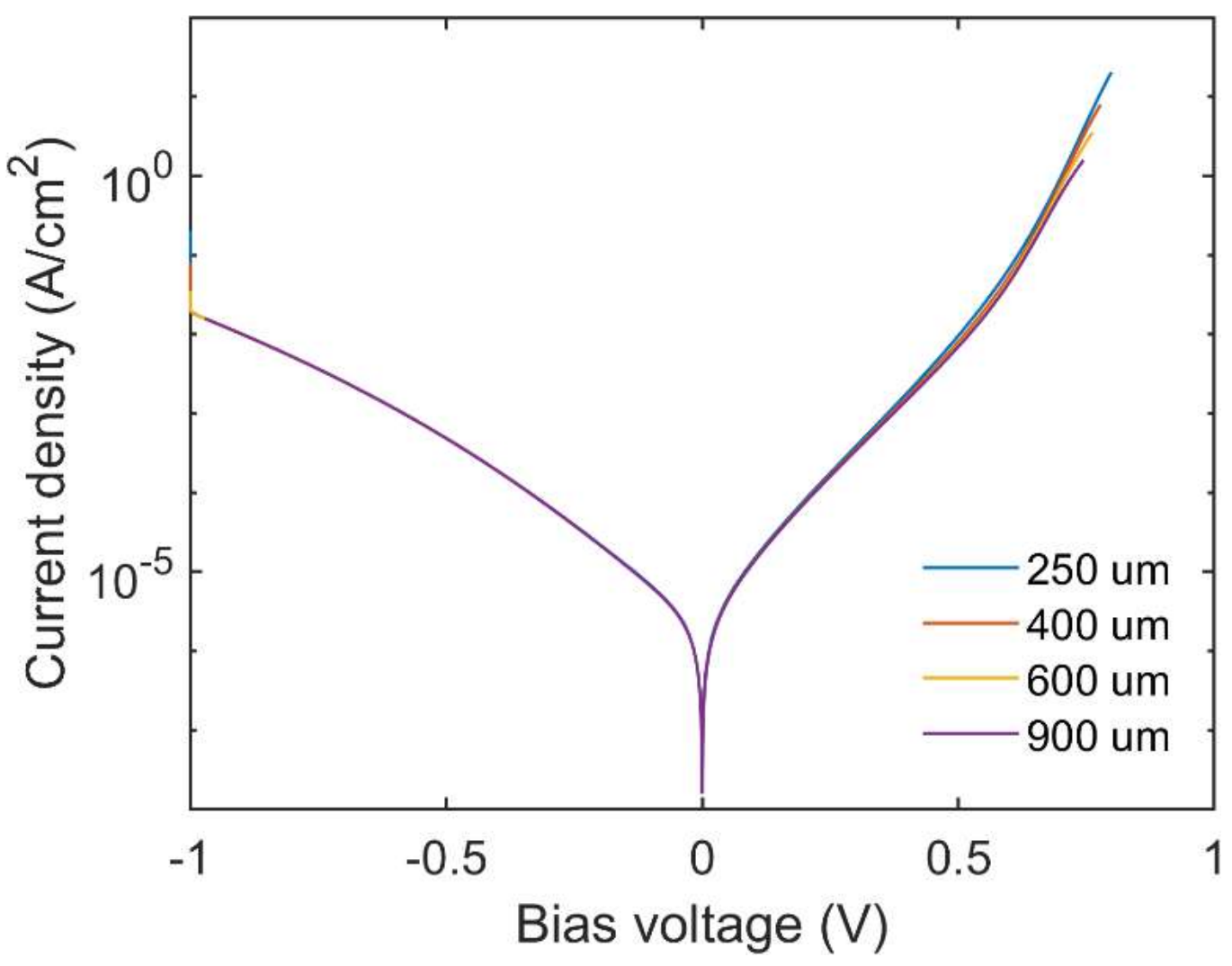

**Figure S2 J-V characteristics obtained from devices with different diameters. The current scales well with device area. The measurements were done under room temperature.**

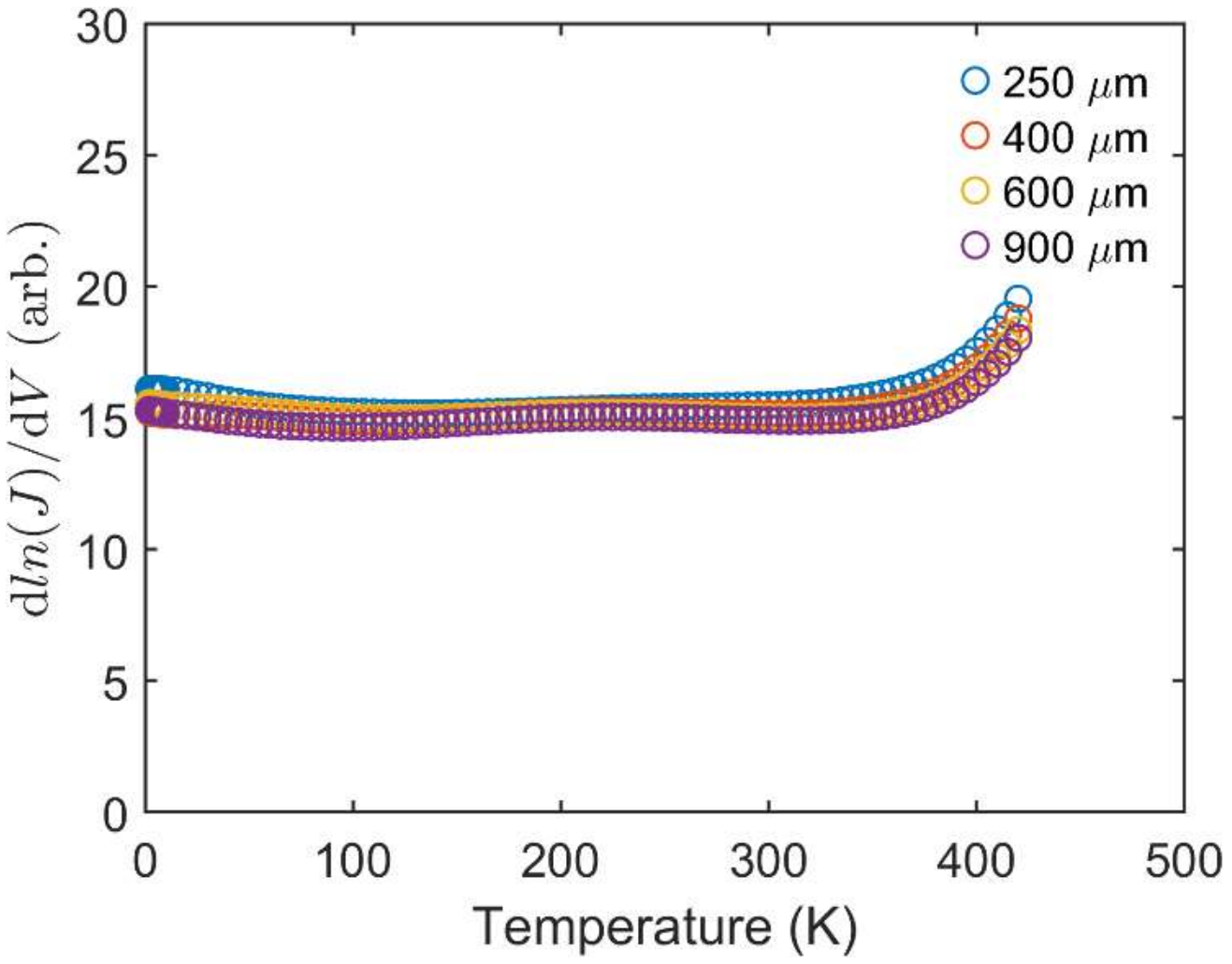

**Figure S3 Log slope as a function of temperature extracted from J-V curves from devices with different diameters. Legend presents diameters of the devices.**

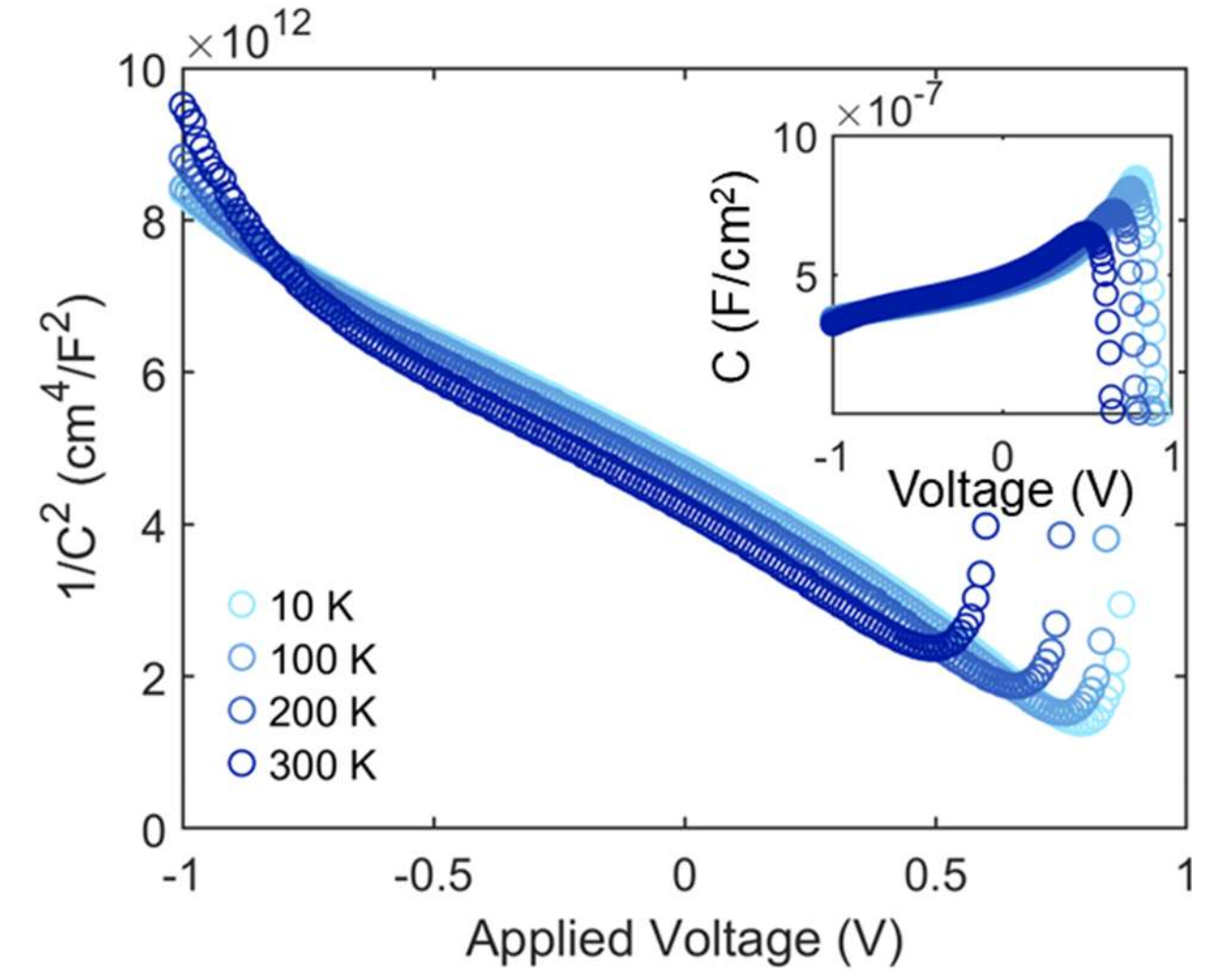

**Figure S4** **1/C²-V curves at various temperatures and corresponding C-V curves (inset).**

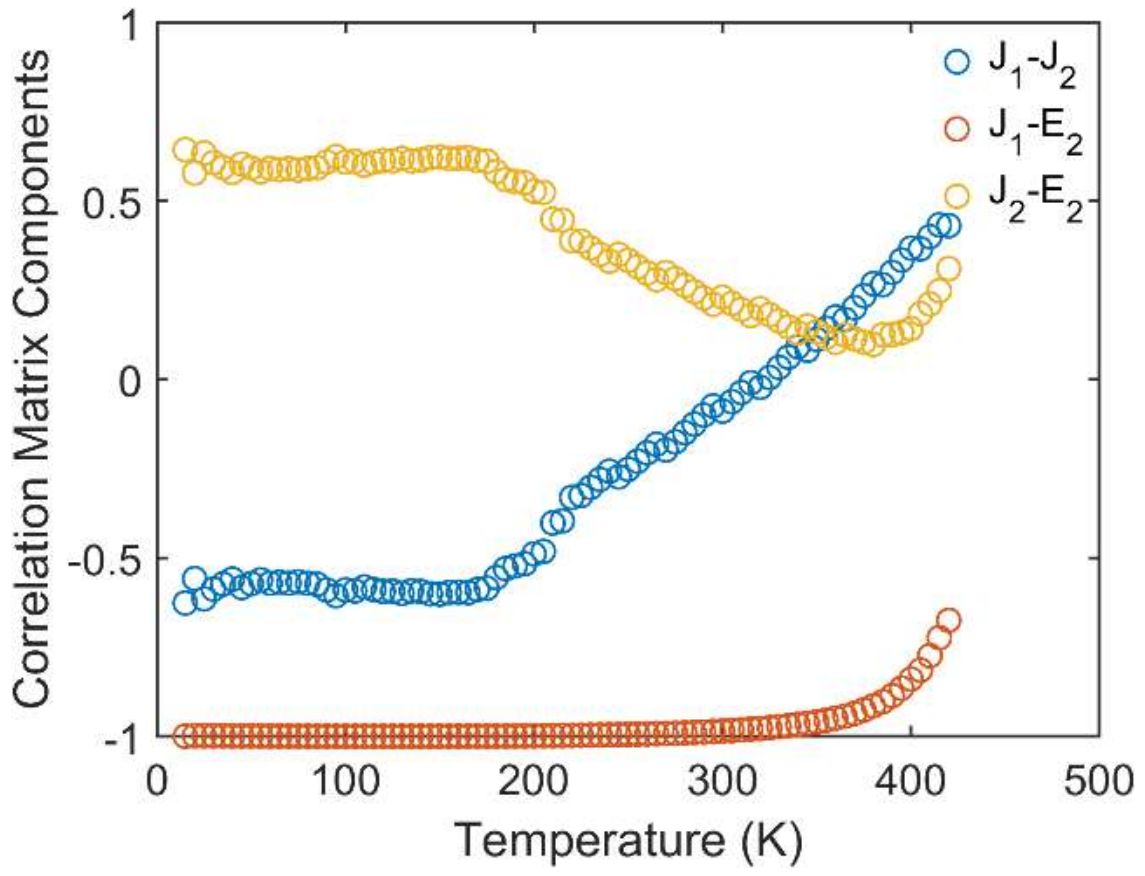

**Figure S5 Correlation matrix component values from the analytic non-linear least square fitting results to J-V characteristics.**

**Supporting Note 1: Correlation between fitting parameters in non-linear least square fitting to J-V characteristics**

Figure S5 presents the correlation matrix components obtained from our analytic fitting to the temperature-dependent J-V characteristics. In non-linear least square fitting, the correlation matrix quantifies the degree of correlation among fitting parameters.

For the calculation of the correlation matrix, the covariance matrix is defined as follows:

$$\boldsymbol{C} = s^2(\boldsymbol{J}^T\boldsymbol{J})^{-1}$$

where $s^2$ is the estimated residual variance and $\boldsymbol{J}$ is the Jacobian matrix evaluated at the optimized parameter values. $s^2$ and $\boldsymbol{J}$ are expressed as follows:

$$s^2 = \frac{(RSS)}{N - p}$$

$$J_{ij} = \partial f(x_i, \beta)/\partial \beta_j$$

where $RSS$ is the residual sum of squares, $N$ is the number of data points, $p$ is the number of fitting parameters, $f(x_i, \beta)$ is the fitted function evaluated at the $i$-th data point $x_i$, and $\beta_j$ is the $j$-th fitting parameter. Finally, the correlation matrix can be obtained as:

$$R_{ij} = \frac{C_{ij}}{\sqrt{C_{ii}C_{jj}}}$$

Although J1 and E2 exhibit noticeable anti-correlation, distinct transport regimes discussed in the main text and the fitting results in Figure 4a justifies the physical validity of the extracted parameters. For the remaining parameter pairs, $J_1$-$J_2$ and $J_2$-$E_2$, they show moderate correlation of $|R_{ij}| < 0.8$.

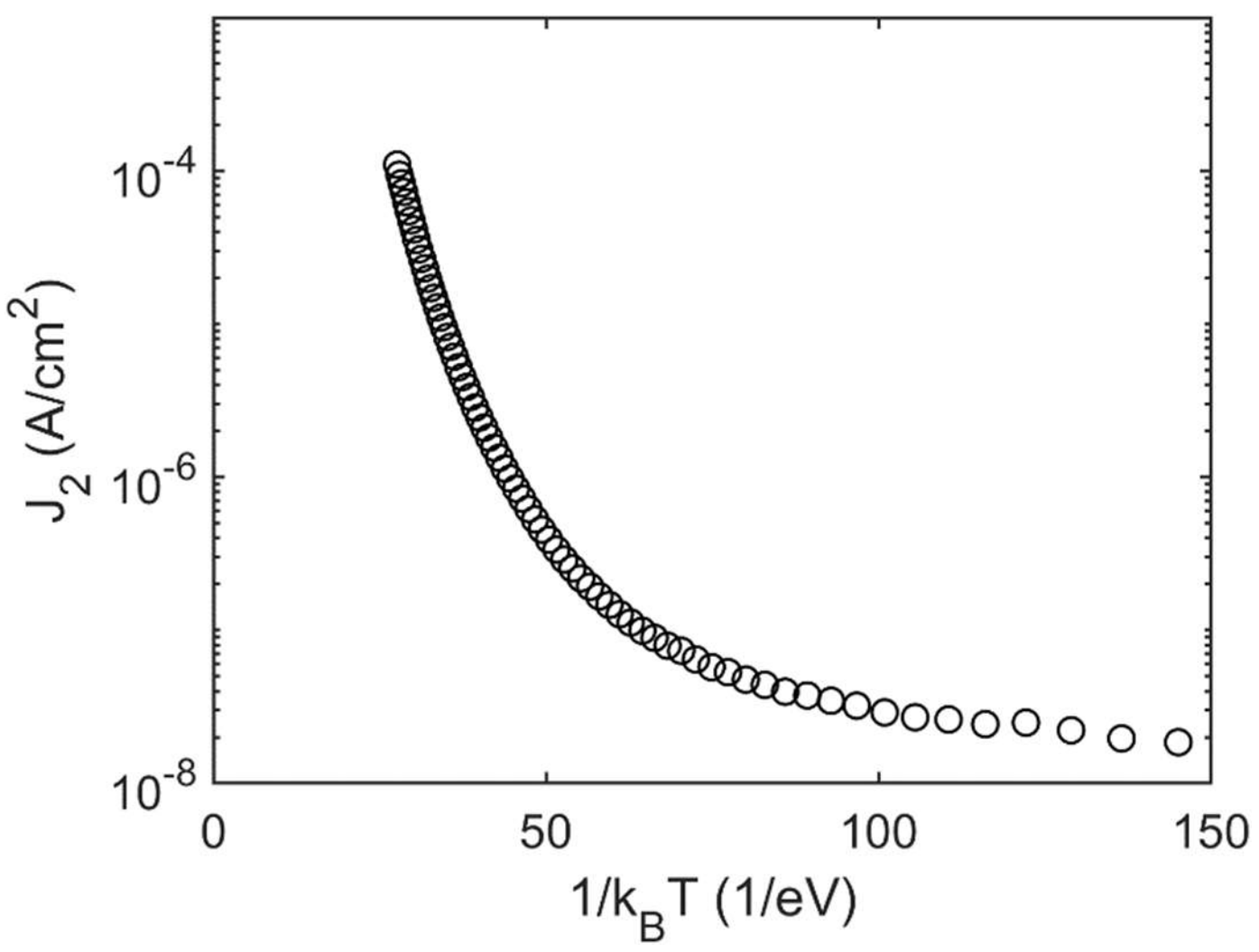

**Figure S6 Arrhenius plot of extracted $J_2$ parameter.**

**Supporting Note 2: Derivarion of the effective multiphonon TAT fitting expression**

We derive a temperature-dependent transition rate using a single-mode multiphonon framework motivated by Makram-Ebeid and Lannoo [1]. The phonon weights follow the displaced harmonic oscillator model. An additional approximation to the electron tunneling kernel yields the final expression used to fit $J_2$. We first describe the lattice fluctuations and then derive the rate from the discrete phonon weights. The connection to the forward bias TAT current prefactor $J_2$ requires further assumptions, which are explained below.

**[Lattice fluctuations and the transition energy]**

Electron emission from an occupied trap to the conduction band changes the equilibrium configuration of the atoms near the defect site. Let's describe this collective displacement by a effective coordinate Q with effective mass M.

Let the displacement for the occupied state to be Q = 0 and for the unoccupied state to be Q = d, respectively. Then, for equal harmonic curvatures, potentials for the occupied state, $U_{occupied}(Q)$, and the unoccupied state $U_{empty}(Q)$, are

$$U_{occupied}(Q) = \frac{1}{2}\kappa Q^2, U_{empty}(Q) = E_t + \frac{1}{2}\kappa(Q-d)^2$$

, respectively, where $\kappa \equiv M\omega^2$ and $E_t$ is the energy between the minimums of the occupied and empty states. $U_{empty}(Q)$ includes the emitted electron at the conduction band edge.

At fixed Q, the energy required to the vertical transition $\Delta(Q)$ is

$$\Delta(Q) = U_{empty}(Q) - U_{occupied}(Q) = E_t - \kappa dQ + \frac{1}{2}\kappa d^2 = E_t - gQ + \lambda$$

, where $g \equiv \kappa d$ and $\lambda \equiv \frac{1}{2}\kappa d^2$.

For the occupied state, the expectation value of the displacement $\langle Q \rangle = 0$, therefore,

$$\bar{\Delta} = E_t + \lambda.$$

For a single phonon mode with energy levels $E_n = \hbar\omega\left(n + \frac{1}{2}\right)$, the expectation value of phonon number $\langle n \rangle = \frac{1}{e^{\hbar\omega/k_B T} - 1}$, and the displacement operator $\hat{Q}$ is

$$\hat{Q} = \sqrt{\frac{\hbar}{2M\omega}}(\hat{a}^\dagger + \hat{a}).$$

Therefore, the variance of Q becomes

$$\sigma_Q^2 = \langle Q^2 \rangle - \langle Q \rangle^2 = \langle Q^2 \rangle - 0 = \frac{\hbar}{2M\omega}\left[\langle \hat{a}^{\dagger 2} \rangle + \langle \hat{a}^{\dagger 2} \rangle + \langle \hat{a}^\dagger \hat{a} \rangle + \langle \hat{a}\hat{a}^\dagger \rangle\right].$$

Since $\langle \hat{a}^{\dagger 2} \rangle = \langle \hat{a}^{\dagger 2} \rangle = 0$ and $\langle \hat{a}^\dagger \hat{a} \rangle = N, \langle \hat{a}\hat{a}^\dagger \rangle = N + 1$ in thermal equilibrium,

$$\begin{aligned}
\sigma_Q^2 &= \frac{\hbar}{2M\omega}(2N + 1) \\
&= \frac{\hbar}{2M\omega}\left(\frac{2}{e^{\hbar\omega/k_B T} - 1} + 1\right) \\
&= \frac{\hbar}{2M\omega}\frac{e^{\hbar\omega/k_B T} + 1}{e^{\hbar\omega/k_B T} - 1} \\
&= \frac{\hbar}{2M\omega}\frac{e^{\hbar\omega/2k_B T} + e^{-\hbar\omega/2k_B T}}{e^{\hbar\omega/2k_B T} - e^{-\hbar\omega/2k_B T}} \\
&= \frac{\hbar}{2M\omega}\coth\frac{\hbar\omega}{2k_B T}.
\end{aligned}$$

Now, we show the variance in the lattice displacement induces the variance in transition energy. The difference of the transition energy at displacement Q from the mean value of the transition energy, $\delta\Delta$ is

$$\delta\Delta = \Delta(\mathrm{Q}) - \bar{\Delta} = -gQ$$

Therefore, the variance in transition energy $\sigma_\Delta^2$ is expressed as

$$\sigma_\Delta^2 = \langle(\delta\Delta)^2\rangle = g^2\langle Q^2\rangle = \frac{\hbar g^2}{2M\omega}\coth\frac{\hbar\omega}{2k_B T}$$

Here, substituting $\kappa = M\omega^2$, $g = \kappa d$, and $\lambda = \frac{1}{2}\kappa d^2$,

$$\sigma_\Delta^2 = \lambda\hbar\omega\coth\frac{\hbar\omega}{2k_B T}$$

This relation explains the effect of the harmonic phonon statistics on the energy fluctuations, which results in the following tunneling rate.

**[Phonon weights and the total tunneling rate]**

As described in S. Makram-Ebeid and M. Lannoo, Physical Review B 25, 6046 (1982), under Condon approximation, the phonon-assisted tunneling rate from trap to band can be written as a weighted sum of electronic tunneling rate, where the elastic electron tunneling rate $\Gamma_{el}$ for each path is weighted by multiphonon weighting function $W_m$, i.e.

$$\Gamma(T) = \sum_{m=-\infty}^{\infty} W_m(T)\Gamma_{el}(\Delta_m)$$

$$W_m = \exp\left(\frac{m\hbar\omega}{2k_BT} - S\coth\frac{\hbar\omega}{2k_BT}\right) I_m\left(\frac{S}{\sinh\frac{\hbar\omega}{2k_BT}}\right).$$

where m, $\Delta_m$, $S$, $I_m$ are the difference of the phonon number before and after the transition, the transition energy corresponding to m-phonon assisted transition, the Huang-Rhys coupling constant, and modified Bessel function of the first kind. Using $\coth\frac{\hbar\omega}{2k_BT} = 2N+1$ and $\sinh\frac{\hbar\omega}{2k_BT} = \frac{1}{2\sqrt{N(N+1)}}$,

$$W_m = \exp\left(\frac{m\hbar\omega}{2k_BT} - S(2N+1)\right) I_m\left(2S\sqrt{N(N+1)}\right)$$

Let's consider a generating function

$$M(u) \equiv \sum_{m=-\infty}^{\infty} W_m(T) e^{um}$$

Using the Bessel identity $\sum_{m=-\infty}^{\infty} I_m(z)t^m = exp\left[\frac{z}{2}(t+t^{-1})\right]$ and the Bose occupation relation $e^{\frac{\hbar\omega}{2k_BT}} = \sqrt{\frac{N+1}{N}}$,

$$M(u) = e^{-S(2N+1)} \sum_{m=-\infty}^{\infty} \left(\sqrt{\frac{N+1}{N}} e^u\right)^m I_m\left(2S\sqrt{N(N+1)}\right)$$

$$= \exp[-S(2N+1)] \exp\left[S\sqrt{N(N+1)}\left(\sqrt{\frac{N+1}{N}} e^u + \sqrt{\frac{N}{N+1}} e^{-u}\right)\right]$$

$$= \exp\left[-S(2N+1) + S\left((N+1)e^u + Ne^{-u}\right)\right]$$

$$= \exp[S\{(N+1)(e^{u}-1)+N(e^{-u}-1)\}].$$

To approximate the electronic tunneling rate for each path, consider a triangular barrier $U(x) = \Delta - qEx$, where $E > 0$ is the electric field magnitude and q is the elementary charge. The WKB expression for the tunneling rate is

$$\Gamma_{el} \propto exp\left(-2\int_0^{x_t} \frac{\sqrt{2m^*(\Delta - qEx)}}{\hbar} dx\right) = exp\left(-\frac{4\sqrt{2mm^*}}{3q\hbar E}\Delta^{3/2}\right)$$

where $x_t = \frac{\Delta}{qE}$ is the classical turning point. Here, $m^*$ is the effective tunneling mass. Taylor expansion around $\overline{\Delta}$ gives

$$\ln \Gamma_{el}(\Delta) \approx \ln \Gamma_{el}(\overline{\Delta}) - \eta(\Delta - \overline{\Delta})$$

where

$$\eta = \frac{2\sqrt{2m^*\overline{\Delta}}}{q\hbar E}$$

Therefore,

$$\Gamma_{el}(\Delta) = \Gamma_{el}(\overline{\Delta})e^{-\eta(\Delta-\overline{\Delta})}$$

So, the tunneling rate depends exponentially on the transition energy near the mean transition energy.

Then, for the dominant multiphonon process paths,

$$\Gamma_{el}(E_t + m\hbar\omega) \approx \Gamma_{el}(E_t)e^{-m\eta\hbar\omega}$$

Therefore,

$$\Gamma(T) \approx \sum_{m=-\infty}^{\infty} W_m(T)\Gamma_{el}(E_t)e^{-m\eta\hbar\omega} = \Gamma_{el}(E_t)M(-\eta\hbar\omega)$$

Defining $\xi \equiv \eta\hbar\omega$,

$$\Gamma(T) \approx \Gamma_{el}(E_t)M(-\xi) = \Gamma_{el}(E_t)\exp\left[S\left\{(N+1)\left(e^{-\xi}-1\right) + N(e^{+\xi}-1)\right\}\right]$$

And since $N = 0$ at T=0,

$$\Gamma(0) \approx \Gamma_{el}(E_t)\exp\left[S\left(e^{-\xi}-1\right)\right]$$

Dividing two results yields,

$$\frac{\Gamma(T)}{\Gamma(0)} \approx \exp\left[SN\left(e^{\xi}+e^{-\xi}-2\right)\right] = \exp[2SN(\cosh\xi - 1)]$$

Using $2N = \coth\frac{\hbar\omega}{2k_BT} - 1$,

$$\Gamma(T) \approx \Gamma(0)\exp\left[S(\cosh\xi - 1)\left(\coth\frac{\hbar\omega}{2k_BT} - 1\right)\right].$$

**[Application to the forward-current prefactor]**

The derivation above describes emission from an occupied trap to the conduction band. A forward bias TAT current in our $p^+n^+$ junction diode additionally involves carrier capture, trap occupancy, and transitions involving the valence band, as considered in Schenk's work.[2] Consequently, the emission rate cannot be identified directly with $J_2$ without further assumption.

We use the normalized temperature dependence from the result above as an effective fitting form,

$$J_2(T) = J_{20} \exp\left[B\left(\coth\frac{\hbar\omega}{2k_B T} - 1\right)\right]$$

This assumes that a rate-limiting transition in the TAT process is the emission from the trap to the conduction band and the other prefactors introduce comparatively weak temperature dependence. The agreement between the experimental data and the model is therefore consistent with a multiphonon-assisted TAT contribution, but does not uniquely establish this mechanism. Here, B combines electron-phonon coupling and the energy sensitivity of the electron tunneling rate. $\hbar\omega$ is an effective vibrational energy. The fitted energy does not represent a specific phonon branch. This expression provides an effective description based on the phonon statistics.